\documentclass[12pt,a4paper]{article}

\usepackage{graphicx,amsmath,epsfig,latexsym,amssymb,color,a4wide,amsfonts,amsbsy}
\usepackage{color}
\usepackage[latin1]{inputenc}

\newcommand{\be}{\begin{equation}}
\newcommand{\ee}{\end{equation}}
\newcommand{\beann}{\begin{eqnarray*}}
\newcommand{\eeann}{\end{eqnarray*}}
\newcommand{\bea}{\begin{eqnarray}}
\newcommand{\eea}{\end{eqnarray}}

\newcommand{\bdm}{\begin{displaymath}}
\newcommand{\edm}{\end{displaymath}}

\begin{document}


\title{A note on testing pre-inflationary times with gravitational-waves}

\author{%
Carlos Marques Afonso\thanks{\textsf{cmmafonso@gmail.com}}\\
Departamento de Fisica and CENTRA, \\ Instituto Superior T\'ecnico,\\  Rua Rovisco Pais, 1049 Lisboa Codex, Portugal.\\
[3mm]
Alfredo Barbosa Henriques\thanks{\textsf{alfredo@fisica.ist.utl.pt}} \\
Departamento de Fisica and CENTRA, \\ Instituto Superior T\'ecnico,\\  Rua Rovisco Pais, 1049 Lisboa Codex, Portugal.\\
[3mm]
Paulo Vargas Moniz\thanks{\textsf{pmoniz@ubi.pt}}~\thanks{%
URL: \textsf{http://www.dfis.ubi.pt/$\sim$pmoniz}}~\thanks{%
Also at CENTRA, IST, Rua Rovisco Pais, 1049 Lisboa Codex, Portugal.}\\
Departamento de F\'{\i}sica, \\ Faculdade de Ci\^encias \\ Universidade da Beira Interior, \\
Rua Marqu\^es d'Avila e Bolama, 6200 Covilh\~a, Portugal.}
\date{}
\maketitle

\begin{abstract}
A simple inflationary model based on loop quantum cosmology is
considered. Within this framework, we show that inflation does not
necessarily erase the information prior to its onset, but that such
information may leave its imprint in the energy-spectrum of the
gravitational-waves generated at these earliest of times.

\end{abstract}





\indent

It is usually assumed that inflation, by its characteristics, among
them a tremendous increase in the scale of the universe taking place
in an extremely short time, will effectively remove any kind of
information or properties already existing prior to the onset of
this phenomenon (cf., e.g., \cite{b:L-L-00} and references therein).
In this work we use a simple example, from the literature on loop
quantum cosmology (LQC) \cite{p:MB-08}-\cite{BojoHoss}, to show that
this is not necessarily the case\footnote{Recently, a few papers
addressing this or other related issues and frameworks, have
appeared in the literature \cite{morenew}}. In particular, from the
model herein analysed, although the very first quantum stages in LQC
are followed by inflation, still, traces of that pre-inflationary
phase survive, leaving their imprint in today's energy-spectrum of
the stochastic background of gravitational waves.

Before proceeding into more technical matters, let us specify
several elements that distinguish  our framework. To calculate the
power spectrum of the gravitational waves, we apply a method based
on the Bogoliubov coefficients and the differential equations obeyed
by them, a method first devised by L. Parker \cite{p:LP-69} and
subsequently used  in other papers
\cite{t:LM-97,p:ABH-04,p:M-H-M-94,Allen}. Furthermore, our model
includes several stages of expansion: a pre-inflationary era,
followed by inflation and then the standard stage, bringing us to
the present universe, whose dark energy component is, for
simplicity, represented by a cosmological constant. In addition, we
avoid using the sudden transition approximation \cite{Allen}  and,
instead, with the help of Parker's differential equations, consider
the transitions between the different stages as continuous.

In the model that we are considering, the potential
$U=\frac{a^{\prime\prime}}{a}$, appearing in the equations for the
gravitational waves ($a$ designates the scale of the universe and
the prime its derivative with respect to conformal time), shows, in
the pre-inflationary phase, a specific feature, not present in the
usual model based on the classical equation for the inflaton field.
This feature is a small peak in the graph of $U$ with respect to
conformal time. This small barrier gives rise to a different set of
initial conditions, for those gravitons whose frequencies satisfy
$k^2<U$. This, in turn, will be reflected in the shape of the
energy-spectrum in the low-frequency range, for frequencies
approximately below $10^{-13}$ rad/s. For gravitons with frequencies
above this, corresponding to energies above $U$, the spectrum will
not be changed, when compared to the classical case.




\indent

We begin with a description of our background model, closely
following \cite{p:T-S-M-03} and \cite{Bojowald:2003uh}. In LQC, a
scalar field $\phi $ with potential $V(\phi )$, in a flat
Friedmann-Robertson-Walker background, is described by the
Hamiltonian \cite{p:T-S-M-03} \be \mathcal{H}_{\phi }=a^{3}V(\phi
)+\frac{1}{2}d_{j,l}p_{\phi }^{2}, \label{1} \ee where $p_{\phi
}=d_{j,l}^{-1}\stackrel{.}{\phi}$
 is the
momentum canonically conjugate to $\phi$   and
$d_{j,l}(a)=D_{l}(q)/a^{3}$ gives the eigenvalues  of the
geometrical density operator in loop quantization; $j$ and $l$ are
the so-called ambiguity parameters, $j$ being a half integer quantum
number greater than one and where $l$ determines the behavior of the
density operator on small scales (compared to a fundamental scale
$a_{*}$, introduced below) and can take any value between zero and
one. $D_{l}(q)$ is a quantum correction factor for the density in
the semiclassical regime \cite{b:CK-04,p:MB-08}. Throughout the
paper we take $j$ as a free parameter. On the other hand, we fix
$l=3/4$. The expression for $D_{3/4}(q)$ is then given by \be
D_{3/4}(q)=\left(\frac{8}{77}\right)^{6}q^{3/2}%
\{7[(q+1)^{11/4}-|q-1|^{11/4}]-11q[(q+1)^{7/4}-sgn(q-1)|q-1|^{7/4}]\}^{6},
\label{2} \ee where the variable $q\equiv(a/a_{*})^{2}$ and
$a_{*}\equiv\sqrt{(\gamma j)/3}l_{Pl}$ is a fundamental length scale
arising in loop quantization; $\gamma $ is the Barbero-Immirzi
parameter, for which calculations of the black-hole entropy give
$\gamma \simeq 0.274$ \cite{b:CK-04,p:MB-08}.

The LQC corrections will appear in the background equations of our
model through the function $D(q)$ and its derivative
$\stackrel{.}{D}(q)$. These effects are particularly relevant for
$q\leq 1$, when the scale factor $a\leq a_{*}$.
For $q\gg 1$ we enter the classical regime, characterized by $%
D(q)\rightarrow 1$ and $\stackrel{.}{D}(q)\rightarrow 0.$ At this
point the semiclassical regime reduces to the classical case.

We assume a universe described by a flat FRW metric and going
through three stages of expansion:

\begin{description}
\item[(a) The pre-inflationary and the inflationary stages:]
To model it, we use the chaotic inflaton potential \be
V(\phi )=\frac{1}{2}m_{\phi }^{2}\phi ^{2}. \label{01} \ee

The background equations are the Friedmann equation,
$H=\stackrel{.}{a}/a$, \be
H^{2}=\frac{8\pi }{3}l_{pl}^{2}\left[V(\phi )+\frac{1}{2D}\stackrel{.}{\phi }%
^{2}\right] \label{3}  \ee
and the modified Klein-Gordon equation for the inflaton, obtained from $%
\mathcal{H}_{\phi }$, \be
\stackrel{..}{\phi }+\left(3\frac{\stackrel{.}{a}}{a}-\frac{\stackrel{.}{D}}{D}\right)%
\stackrel{.}{\phi }+DV_{\phi }(\phi )=0, \label{4}  \ee where we
take derivatives with respect to comoving time. The classical
equations are obtained putting $D=1$ and $\stackrel{.}{D}=0$.

\item[(b) Transition between inflation and the radiation dominated universe:]
We define the end of the inflationary era, beginning of the
transition era, as the point where $\stackrel{..}{a}$ becomes
negative. During the transition, the energy of the inflaton will be
transferred to the radiation fluid. We model this transfer through
the action of a frictional term characterized by a decay constant
$\Gamma _{\phi }$. The equations above are now slightly changed. The
Friedmann equation becomes \cite{p:M-M-S-93} \be
H^{2}=\frac{8\pi }{3}l_{pl}^{2}\left[V(\phi )+\frac{1}{2}\stackrel{.}{\phi }%
^{2}+\rho _{R}\right] \label{6}  \ee and the equation for the scalar
field is \be \stackrel{..}{\phi
}+3\frac{\stackrel{.}{a}}{a}\stackrel{.}{\phi }+V_{\phi }(\phi
)=-\Gamma _{\phi }\stackrel{.}{\phi }. \label{7} \ee The equation
for the radiation fluid is (we use $\Gamma_\phi \simeq 2 \times
10^{-7}$): \be
\stackrel{.}{\rho }_{R}=-4\frac{\stackrel{.}{a}}{a}\rho _{R}+\Gamma _{\phi }%
\stackrel{.}{\phi }^{2}. \label{8}  \ee At the beginning of the
transition era we assume $\rho_R = 0$. We end this period when when
$\rho _{R} a^{4} = const. =\rho _{R0} a_{0}^{4}$.

\item[(c) Standard stage:] This brings us to the present universe, whose dark energy component is
taken, for simplicity, to be   described by a cosmological constant.
We, then, integrate the equations till the present time, with the
content of the universe characterized by a radiation field $\rho
_{R}$, dark matter+baryonic matter, with equation of state $p=0$,
and a cosmological constant representing the dark energy. The
Friedmann equation is now \be
H^{2}=\frac{8\pi G}{3} \left[\rho _{R0}\left(\frac{a_{0}}{a}\right)^{4}+\rho _{M0}\left(\frac{a_{0}%
}{a}\right)^{3}+\rho _{DE0}\right], \label{9}  \ee the subscript $0$
denoting the values at the present time, $a(t_{0})=a_{0}$.

\end{description}



Above, we wrote our expressions using comoving time (as this is the
most appropriate variable when we have the fast increase in the
scale factor during inflation). However, the equations for the
gravitational waves take a much simpler form when we use conformal
time $\tau $, defined by writing the FRW metric in the form \be
ds^{2}=s^{2}(\tau )\left\{ -d\tau ^{2}+\left[ \delta _{ij}+h_{ij}(\tau ,%
\mathbf{x})\right] dx^{i}dx^{j}\right\} . \label{10}  \ee

The tensor perturbations $h_{ij}$ can be expanded, in the usual
manner, in terms of plane waves: \be h_{ij}(\tau
,\mathbf{x})=\sqrt{8\pi G}\sum_{p=1}^{2}\int \frac{d^{3}k}{(2\pi
)^{3/2}a(\tau )\sqrt{2k}}\left[ a_{p}(\tau ,\mathbf{k})\varepsilon _{ij}(%
\mathbf{k},p)e^{i\mathbf{k}\cdot \mathbf{x}}\xi (\tau
,k)+\text{h.c.}\right] , \label{11}  \ee
 where $\mathbf{x}$ denotes
the spatial coordinates, $k=|\mathbf{k}|=2\pi /\lambda =\omega a$ is
the comoving wave number; the index $p$ runs over the two
polarizations of the gravitational waves and $\varepsilon _{ij}$ is
the polarization tensor, $a_{p}$ the annihilation operator and $\xi
$ the mode function. In conformal time, the mode function obeys the
equation
 \be \xi ^{\prime \prime }+(k^{2}-a^{\prime \prime }/a)\xi
=0, \label{12}  \ee
 the derivatives being with respect
to
 $\tau $. In what
follows, we call gravitational wave potential the expression \be
U=\frac{a^{\prime \prime }}{a}. \label{27}  \ee This potential is
shown in figure 1, for the semiclassical case (cf. eq.
(\ref{01})-(\ref{4})). The LQC main feature appears as an additional
small barrier at very early times, \emph{prior to the onset of the
inflationary stage}. This small barrier gives rise to different
initial conditions at the end of the pre-inflationary era, for those
gravitons whose frequencies $k^2<U$. This, in turn, will be
reflected in the shape of the energy-spectrum in the low-frequency
range, as will be seen.

\begin{figure}[!ht]
   \centering
        \includegraphics[width=0.8\textwidth]{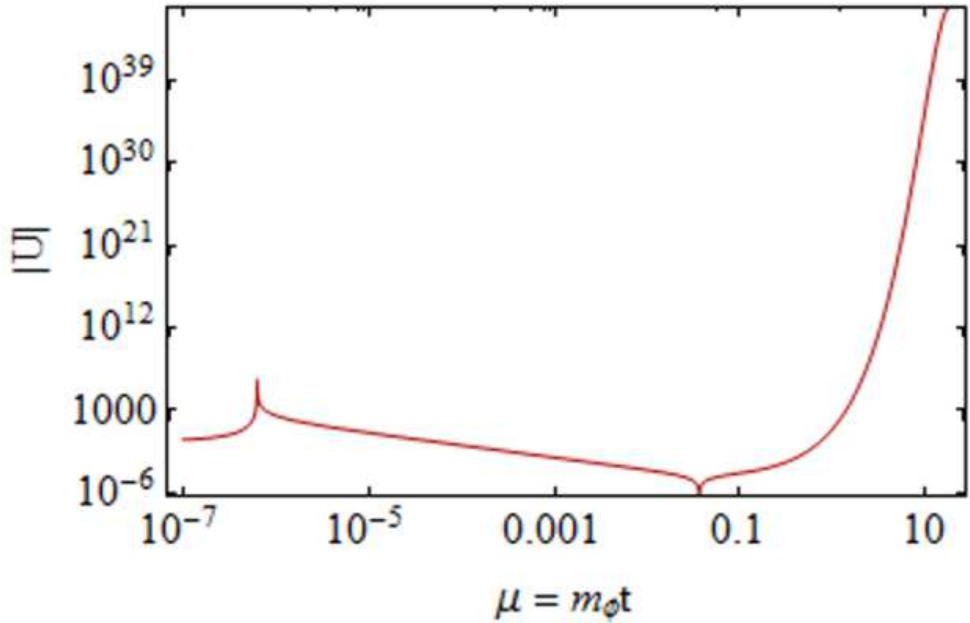}
    \caption{Gravitational wave potential $|U|$, as it appears in the LQC model herein employed.}
    \label{fig1}
\end{figure}

More generically, for appropriate forms of $a''(\tau)/a(\tau)$  we
may have exponentially growing solutions, with the gravitational
field pumping energy into the gravitational waves. From the point of
view of quantum mechanics, this corresponds to graviton production.
In fact, equation (\ref{12}) closely mimics a Schrödinger equation
with potential barrier $U(\tau) = a''(\tau)/a(\tau)$ (with the time
coordinate $\tau$ instead of the spatial coordinate). It is an
important point to realize that, whenever $k^{2}\lesssim U(\tau )$,
the conditions exist for a significant graviton production, the
equation for the gravitational waves being then of the form of a
parametric oscillator, while, when $k^{2}\gg U(\tau )$, the equation
for the gravitational waves is that of a harmonic oscillator, with
no gravitons being produced.

We now express the creation and annihilation operators in terms of
the initial creation and annihilation operators $A_{p}^{\dagger
}(\mathbf{k})$ and $A_{p}(\mathbf{k})$, through the Bogoliubov
coefficients $\alpha (k,\tau )$ and $\beta (k,\tau )$
\cite{p:LP-69}, \cite{b:M-W-07}: \be a_{p}(\tau ,\mathbf{k})=\alpha
(\tau ,k)A_{p}(\mathbf{k})+\beta ^{*}(\tau ,k)A_{p}^{\dag
}(\mathbf{k}), \label{13}  \ee $\alpha $ and $\beta $ satisfying the
relation \be |\alpha |^{2}-|\beta |^{2}=1. \label{14}  \ee These
coefficients, which in the sudden transition approximation (cf.
\cite{Allen,p:MG-99})  are calculated by requiring the mode
functions and their derivatives to be continuous across the
transition, obey the following couple of differential equations
\cite{p:LP-69} \be \alpha ^{\prime }=\frac{i}{2k}\left[ \alpha
+\beta e^{2ik(\tau -\tau _{i})}\right] \frac{a^{\prime \prime }}{a}
\label{15}  \ee and \be \beta ^{\prime }=-\frac{i}{2k}\left[ \beta
+\alpha e^{-2ik(\tau -\tau _{i})}\right] \frac{a^{\prime \prime
}}{a}, \label{16}  \ee which, with a change to the variables
$X(k,\tau )$ and $Y(k,\tau )$ \be \alpha =\frac{1}{2}(X+Y)e^{ik(\tau
-\tau _{i})} \label{17}  \ee \be \beta =\frac{1}{2}(X-Y)e^{-ik(\tau
-\tau _{i})}, \label{18} \ee become \be X^{\prime }=-ikY \label{19}
\ee \be Y^{\prime }=-\frac{i}{k}\left( k^{2}-\frac{a^{\prime \prime
}}{a}\right) X, \label{20}  \ee completing our system of equations.
These variables obey the constraint equation (\ref{14}), which takes
the form \be X_{r}Y_{r}+X_{i}Y_{i}=1, \label{21}  \ee the subscripts
$r$ and $i$ denoting the real and imaginary parts. In terms of the
new variables, $|\beta |^{2}$ is \be |\beta |^{2}=\frac{1}{4}\left[
(X_{r}-Y_{r})^{2}+(X_{i}-Y_{i})^{2}\right] . \label{22}  \ee

During  the LQC dominated phase, we have, as mentioned,  an extra
barrier (cf. Fig. 1). We verified numerically that the appropriate
equation of state is $p_{\phi }\simeq \rho _{\phi }$, in which case
an analytical solution of equations (\ref{19}) and (\ref{20}) can be
found and is of the form \be X=\sqrt{\frac{\pi \theta }{2}}\left[
-J_{0}(\theta )+iY_{0}(\theta )\right], \label{32} \ee \be
Y=\sqrt{\frac{\pi }{8\theta }}\left\{ -Y_{0}(\theta )+2\theta
Y_{1}(\theta )+i\left[ J_{0}(\theta )-2\theta J_{1}(\theta )\right]
\right\}, \label{33} \ee where $J_{n}$ and $Y_{n}$ are the Bessel
functions of the first and second kind, respectively. In this case,
$|\beta |^{2}$ has a more complicated form,
but it also satisfies the limits $|\beta |^{2}\rightarrow 0$, when $%
k\rightarrow \infty $, and $|\beta |^{2}\rightarrow \infty $, when $%
k\rightarrow 0$. Due to the smallness of the extra barrier, only the
low frequency gravitons will be affected. For the higher frequency
modes, only the inflationary barrier will be important.

Initial conditions for the scale factor $a_i$ and for $\dot{\phi}_i$
need to be set, with $\phi_i$ being obtained from $\dot{\phi}_i$
with the help of the uncertainty principle (see \cite{p:T-S-M-03})
\be
|\phi _{i}\stackrel{.}{\phi }_{i}|\gtrsim \frac{10^{3}}{j^{3/2}}\left(\frac{a_{i}%
}{a_{*}}\right)^{12}m_{Pl}^{3}. \label{5}  \ee The scale factor is
$a_i \geq  l_{Pl}$, otherwise the semiclassical equations would not
be valid, due to the discretization of the geometry of space
introduced by LQG. We therefore take $a_i = \sqrt{\gamma} l_{Pl}$.

We choose parameters and initial conditions such that we get enough
inflation (number of e-folds $\simeq 70$) and satisfy the
large-scale CMB anisotropies requirements \cite{b:L-L-00}: $m_\phi
\simeq 10^{-6} m_P$ and $\phi_{max} \gtrsim 3 m_P$.
We still have the ambiguity parameter $j$, as a free parameter.

In the inflationary era we used, for convenience, Planck units and
worked with the adimensional time $\mu = m_\phi t$. Finally, we used
the same values for those parameters which are common to both
models, $m_\phi = 10^{-6}$, $\Gamma_\phi = 2\times 10^{-7}$
and $\dot{\phi}_i / m_\phi = 2$ for the semiclassical
and $\phi_i = 3.3$ for the classical
cases.

It is now possible to integrate the full set of equations, including
the equations for $X$ and $Y$, (\ref{19}) and (\ref{20}). The number
of gravitons of gravitons created is, as is well known, given
\cite{p:LP-69,p:ABH-04,Allen}
 by $|\beta _{final}|^{2}$. Taking into
account that the density of states is $\omega ^{2}d\omega /(2\pi
^{2}c^{3})$ and that each graviton contributes with two
polarizations $2\hbar \omega $, then, from the definition of the
energy density $dE=P(\omega )d\omega $, we have the following
expression for the power-spectrum $P(\omega )$: \be P(\omega
)=\frac{\hbar \omega ^{3}}{\pi ^{2}c^{3}}|\beta _{final}|^{2}.
\label{23}  \ee We  choose to  present our results in terms of the
dimensionless relative
logarithmic energy-spectrum of the gravitational waves, at present time $%
\tau _{0}:$ \be \Omega (\omega _{0},\tau _{0})=\frac{1}{\rho
_{crit}(\tau _{0})}\frac{d\rho
_{gw}}{d\ln \omega }(\tau _{0})=\frac{8\hbar G}{3\pi c^{5}H^{2}(\tau _{0})}%
\omega _{0}^{4}|\beta _{final}|^{2}, \label{24}  \ee
$\rho _{crit}$ being the value of the present time critical density and $%
\rho _{gw}$ the gravitational wave energy density \be \rho
_{gw}=\int P(\omega )d\omega. \label{25}  \ee

\begin{figure}[!ht]
   \centering
        \includegraphics*[0cm,15cm][12cm,20cm]{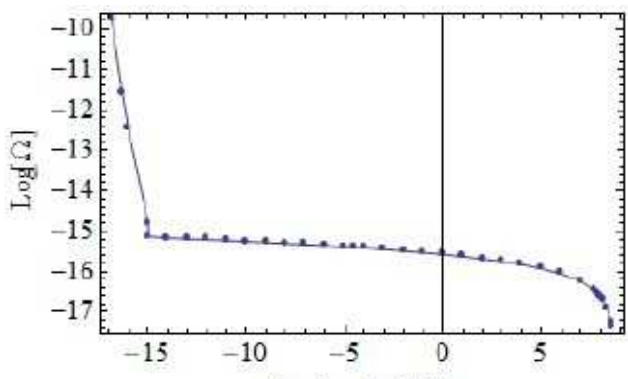}
    \caption{The energy-spectrum of the gravitational waves, for
    the LQC model employed in this paper, showing the sharp
    decrease in the low-frequency part of the spectrum.}
    \label{fig2}
\end{figure}

Our results for the energy-spectrum of the gravitational waves are
plotted in figure 2.
 For
purposes of illustration, we defined values for the parameters of
the classical model such that the results would approximately
comply with the total amount of anisotropy measured by COBE which,
in terms of
%
 $\Omega $, is \be \Omega
\lesssim 1.37\times 10^{-10}, \label{26} \ee
 for
frequencies corresponding to the present horizon size $\omega
_{hor}$, where we use $H=71\,km\,s^{-1}$ $Mpc^{-1}$, giving $\omega
_{hor} \simeq 10^{-17}\,rad\,s^{-1}$. We immediately notice that the
gradient of the low-frequency part of the spectrum is much steeper
than in the classical case, a change that is due to the extra
barrier in Figure 1. This shows that physical features present in
the pre-inflationary eras are not erased out by inflation, as is
commonly assumed, but may leave their imprint in the spectrum of the
gravitational waves\footnote{Recently, a range of publications
focusing on gravitational waves within LQC have appeared in the
literature \cite{new1}-\cite{newlastf}. However, neither of them has
pointed and addressed the issue of physical features from a
pre-inflationary epoch being still extractable in observations. Most
deal with inflation in LQC and the production and propagation of
gravitational waves \emph{during} inflation.}, which was the main
point of our brief note. We leave to a future paper a more thorough
investigation of this topic.





\section*{Acknowledgements}
This work was supported by  POCI/FIS/57547/2005. The authors are
grateful to M. Bojowald for conversations and correspondence.

\end{document}